\documentstyle[12pt,epsfig]{article}
\newcommand{\bea}{\begin{eqnarray}}
\newcommand{\eea}{\end{eqnarray}}
\newcommand{\be}{\begin{equation}}
\newcommand{\ee}{\end{equation}}
\newcommand{\nn}{\nonumber}

\begin{document}

\begin{center}
{\LARGE\bf On the nature of $f_0$ mesons below 1.9 GeV and lower scalar nonets}
\end{center}
\bigskip

\begin{center}  {\large\bf Yu.S.~Surovtsev}
\vspace*{0.5mm}
\end{center}

\begin{center}{\it Bogoliubov Laboratory of Theoretical Physics, JINR,
Dubna 141 980, Russia}
\end{center}

\begin{center}  {\large\bf D.~Krupa} and {\large\bf M.~Nagy}
\vspace*{0.5mm}
\end{center}

\begin{center}{\it Institute of Physics, SAS, D\'ubravsk\'a cesta 9,
842 28 Bratislava, Slovakia}
\end{center}

\begin{abstract}
The combined 3-channel analysis of experimental data on the coupled
processes $\pi\pi\to\pi\pi,K\overline{K},\eta\eta$ is carried out in the
channel with the vacuum quantum numbers. An approach, using only first
principles (analyticity and unitarity) and the uniformizing variable, is
applied.
Definite indications of the QCD nature of the $f_0$ resonances below
1.9 GeV are obtained, among them a surprising indication for $f_0(980)$
to be the $\eta\eta$ bound state. An assignment of the scalar mesons below
1.9 GeV to lower nonets is proposed.
\end{abstract}

{\bf Outline}:
\begin{itemize}
\item Motivation.
\item Three-coupled-channel formalism (resonance pole-clusters, uniformization,\\
the Le Couteur-Newton relations, the background).
\item Combined analysis of experimental data.
\item Lower scalar $0^{++}$ nonets.
\item Conclusions.
\end{itemize}

\section{Motivation}
Already several decades, a problem of scalar mesons draws permanently an
attention of investigators. This is related to an important role played by
these mesons (especially the so-called "$\sigma$-meson") in the hadronic
dynamics. For example, a recent discovery of the $\sigma$-meson below
1 GeV \cite{PDG-02} leads to an important conclusion about the linear
realization of chiral symmetry \cite{SKN-epja02}). Now, phenomena and
quantities are known, an explanation of which is
impossible without the $\sigma$-meson. These are
\begin{itemize}
\item
$a_0^0(\pi\pi)$ -- theories with the nonlinear realization of chiral
symmetry give too small value (see, {e.g.}, \cite{SKN-epja02});
\item
rather big experimental value of the $\pi-N$ sigma term $\Sigma_{\pi N}=
{\hat m}<{\bar u}u+{\bar d}d>$ \cite{Jaffe-Korpa};
\item
the enhancement of the $\Delta I=1/2$ processes in the
$K^0\to\pi^+\pi^-,\pi^0\pi^0$ decays can be nicely accounted for through
the correlation in the scalar channel as summarized by the $\sigma$-meson
\cite{Shabalin};
\item
the phase shift analyses of the $N-N$ scattering in the $^1S_0$ channel
have discovered an attraction in the intermediate range ($1\sim 2$ fm),
which is indispensable for the binding of a nucleus and is stipulated by
the light $\sigma$-meson exchange \cite{Holinde}.
\end{itemize}
In spite of the
above-cited facts and obtained evidences of the existence of the
$\sigma$-meson \cite{SKN-epja02},\cite{Torn}-\cite{Li-Zou-Li}, it seems
in view of the known success of the chiral perturbation theory
(with the nonlinear realization of chiral symmetry) in accounting for
many low-energy phenomena, a number of physicists questions till now
the existence of the $\sigma$-meson (see, {e.g.}, \cite{MO-02}).  Therefore,
let us indicate once more, why we state that in our combined analysis of
the processes $\pi\pi\to\pi\pi,K\overline{K}$ data in the channel with
$I^GJ^{PC}=0^+0^{++}$, a real evidence for the existence of the
$\sigma$-meson has been given \cite{SKN-epja02,PRD-01}:
\begin{itemize}
\item Our approach is rather model-independent because it is based only on
the first principles (analyticity and unitarity) immediately applied to
experimental data analysis, and it is free from dynamical assumptions. At
its realization, we use only the {\it mathematical} fact that a local
behaviour of analytic functions determined on the Riemann surface is
governed by the nearest singularities on all corresponding sheets.
\item In this approach, resonance is represented by the pole cluster
(poles and zeros on the Riemann surface) of the definite type related to
its nature. We have obtained the pole cluster corresponding to the
$\sigma$-meson (the pole position on sheet II is $0.6-i0.605$ GeV).
\item A parameterless description of the $\pi\pi$ background is given only
by allowance for the left-hand branch-point in the uniformizing variable.
This solves an earlier-mentioned problem that the wide-resonance parameters
are largely controlled by the nonresonant background (see, {\it e.g.},
\cite{Achasov-Shest}). Moreover, we have shown that the large background,
obtained in earlier analyses of the $s$-wave $\pi\pi$ scattering, hides,
in reality, the $\sigma$-meson below 1 GeV.
\item The fact, that the parameterless description of the $\pi\pi$
background has been obtained, means that the $\rho$-meson exchange
contribution on the left-hand cut is compensated by the scalar meson
(the $\sigma$-meson) exchange one that has the opposite signs due to gauge
invariance.
\end{itemize}
However, in the $\sigma$-meson pole-cluster, the imaginary part of the
pole on sheet III is too small (the pole cluster must be rather compact
formation). This tells us that it ought to take into account yet an
additional important channel and to consider a 3-channel problem. We
suppose here that this additional channel is the $\eta\eta$ one. It is
clear that this consideration will give additional information about other
$f_0$ mesons.

The $f_0$ mesons are
most direct carriers
of information about the QCD vacuum. The contemporary obscurities in
understanding the scalar sector reflect a level of our knowledge about
the QCD vacuum and about its influence on the hadron spectrum and their
properties. Generally, it seems that the problem of scalar mesons will
be fully solved simultaneously with the solution of the QCD-vacuum one.
Therefore, every step in understanding nature of the $f_0$ mesons is
especially important.

\section{Three-coupled-channel formalism}

We consider the processes $\pi\pi\to\pi\pi,K\overline{K},\eta\eta$ in the
3-channel approach. Therefore, the $S$-matrix is determined on the
8-sheeted Riemann surface. The matrix elements $S_{\alpha\beta}$, where
$\alpha,\beta=1(\pi\pi), 2(K\overline{K}),3(\eta\eta)$, have the
right-hand cuts along the real axis of the $s$ complex plane ($s$ is the
invariant total energy squared), starting with $4m_\pi^2$, $4m_K^2$, and
$4m_\eta^2$, and the left-hand cuts. The Riemann-surface sheets are
numbered according to the signs of analytic continuations of the channel
momenta $$k_1=(s/4-m_\pi^2)^{1/2},~~~~k_2=(s/4-m_K^2)^{1/2}~~~~
k_1=(s/4-m_\pi^2)^{1/2}$$ as follows:
signs $({\mbox{Im}}k_1,{\mbox{Im}}k_2,{\mbox{Im}}k_3)=
+++,-++,--+,+-+,+--,---,-+-, ++-$ correspond to sheets
I, II,$\cdots$, VIII.

The resonance representations on the Riemann surface are obtained with the
help of the formulae (Table 1) \cite{KMS-nc96}, expressing
analytic continuations of the matrix elements to unphysical sheets in
terms of those on sheet I -- $S_{\alpha\beta}^I$ that have only zeros
(beyond the real axis) corresponding to resonances, at least, around the
physical region. In Table 1, the
superscrupt $I$ is omitted to simplify the notation, $\det S$ is the
determinant of the $3\times3$ $S$-matrix on sheet I, $D_{\alpha\beta}$ is
the minor of the element $S_{\alpha\beta}$, that is,
$D_{11}=S_{22}S_{33}-S_{23}^2$, $D_{12}=S_{12}S_{33}-S_{13}S_{23}$, etc.
%\newpage
\begin{center} Table 1.  \end{center}
-------------------------------------------------------------------------------------------------------------------\\
\hspace*{2.8cm}    I \hspace*{1.2cm}  II \hspace*{0.96cm}   III \hspace*{0.89cm}      IV \hspace*{0.95cm}        V \hspace*{1.1cm}       VI \hspace*{1.cm}        VII \hspace*{0.7cm}               VIII     \\
~~~~-------------------------------------------------------------------------------------------------------------------
$$1\rightarrow 1~~~~~~~~   S_{11}~~~~~~~ \frac{1}{S_{11}}~~~~~~~ \frac{S_{22}}{D_{33}}~~~~~~ \frac{D_{33}}{S_{22}}~~~~~~ \frac{\det S}{D_{11}}~~~~~ \frac{D_{11}}{\det S}~~~~~~ \frac{S_{33}}{D_{22}}~~~~~~~ \frac{D_{22}}{S_{33}}$$
$$~1\rightarrow 2~~~~~~~~   S_{12}~~~~~~ \frac{iS_{12}}{S_{11}}~~~~~~\frac{-S_{12}}{D_{33}}~~~~~~\frac{iS_{12}}{S_{22}}~~~~~\frac{iD_{12}}{D_{11}}~~~~~~\frac{-D_{12}}{\det S}~~~~~\frac{iD_{12}}{D_{22}}~~~~~~\frac{D_{12}}{S_{33}}~$$
$$2\rightarrow 2~~~~~~~~   S_{22}~~~~~~~ \frac{D_{33}}{S_{11}}~~~~~~~ \frac{S_{11}}{D_{33}}~~~~~~~\frac{1}{S_{22}}~~~~~~~ \frac{S_{33}}{D_{11}}~~~~~ \frac{D_{22}}{\det S} ~~~~~\frac{\det S}{D_{22}}~~~~~~\frac{D_{11}}{S_{33}}$$
$$~~1\rightarrow 3~~~~~~~~   S_{13}~~~~~~ \frac{iS_{13}}{S_{11}}~~~~~~\frac{-iD_{13}}{D_{33}}~~~~\frac{-D_{13}}{S_{22}}~~~~\frac{-iD_{13}}{D_{11}}~~~~\frac{D_{13}}{\det S}~~~~~\frac{-S_{13}}{D_{22}}~~~~~~\frac{iS_{13}}{S_{33}}~$$
$$~~2\rightarrow 3~~~~~~~~   S_{23}~~~~~~ \frac{D_{23}}{S_{11}}~~~~~~~ \frac{iD_{23}}{D_{33}}~~~~~~\frac{iS_{23}}{S_{22}}~~~~~~\frac{-S_{23}}{D_{11}}~~~~~\frac{-D_{23}}{\det S}~~~~~\frac{iD_{23}}{D_{22}}~~~~~~\frac{iS_{23}}{S_{33}}~$$
$$~3\rightarrow 3~~~~~~~~   S_{33}~~~~~~ \frac{D_{22}}{S_{11}}~~~~~~ \frac{\det S}{D_{33}}~~~~~~\frac{D_{11}}{S_{22}}~~~~~~~\frac{S_{22}}{D_{11}}~~~~~~\frac{D_{33}}{\det S}~~~~~~\frac{S_{11}}{D_{22}}~~~~~~~\frac{1}{S_{33}}$$
--------------------------------------------------------------------------------------------------------------------\\

\hspace*{-0.77cm}
 These formulae
immediately give the resonance representation by poles and zeros on the
Riemann surfaces if one starts from resonance zeros on sheet I. Whereas
in the 2-channel approach, we had 3 types of resonances described by a
pair of conjugate zeros on sheet I: ({\bf a}) in $S_{11}$, ({\bf b}) in
$S_{22}$, ({\bf c}) in each of $S_{11}$ and $S_{22}$, in the 3-channel case,
we obtain 7 types of resonances corresponding to conjugate resonance zeros on
sheet I of ({\bf a}) $S_{11}$; ({\bf b}) $S_{22}$; ({\bf c}) $S_{33}$;
({\bf d}) $S_{11}$ and $S_{22}$; ({\bf e}) $S_{22}$ and $S_{33}$; ({\bf f})
$S_{11}$ and $S_{33}$; and ({\bf g}) $S_{11}$, $S_{22}$, and $S_{33}$. For
example, the arrangement of poles corresponding to a ({\bf g}) resonance is:
each sheet II, IV, and VIII contains a pair of conjugate poles at the points
that are zeros on sheet I; each sheet III, V, and VII contains two pairs of
conjugate poles; and sheet VI contains three pairs of poles.

A resonance of every
type is represented by a pair of complex-conjugate clusters (of poles and
zeros on the Riemann surface) of a size typical for strong interactions. The
cluster kind is related to the state nature. The resonance coupled relatively
more strongly to the $\pi\pi$ channel than to the $K\overline{K}$ and
$\eta\eta$ ones is
described by the cluster of type ({\bf a}); if the resonance is coupled
more strongly to the $K\overline{K}$ and $\eta\eta$ channels than to
the $\pi\pi$, it is represented by the
cluster of type ({\bf e}) (say, the state with the dominant $s{\bar s}$
component); the flavour singlet ({\it e.g.}, glueball) must be represented by
the cluster of type ({\bf g}) as a necessary condition for the ideal case,
if this state lies above the thresholds of considered channels.

Furthermore, according to the type of pole clusters, we can distinguish, in a
model-independent way, a bound state of colourless particles ({\it e.g.},
$K\overline{K}$ molecule) and a $q{\bar q}$ bound state \cite{KMS-nc96,MP-93}.
Just as in the 1-channel case, the existence of a particle bound-state means
the presence of a pole on the real axis under the threshold on the physical
sheet, so in the 2-channel case, the existence of a particle bound-state in
channel 2 ($K\overline{K}$ molecule) that, however, can decay into channel
1 ($\pi\pi$ decay), would imply the presence of a pair of complex conjugate
poles on sheet II under the second-channel threshold without an accompaniment
of the corresponding shifted pair of poles on sheet III. Namely, according
to this test, earlier, the interpretation of the $f_0(980)$ state as a
$K\overline{K}$ molecule has been rejected. In the 3-channel case,
the bound-state in channel 3 ($\eta\eta$) that, however, can decay into
channels 1 ($\pi\pi$ decay) and 2 ($K\overline{K}$ decay), is represented by
the pair of complex conjugate poles on sheet II and by a shifted poles on
sheet III under the $\eta\eta$ threshold without an accompaniment of the
corresponding poles on sheets VI and VII.

For the combined analysis of experimental data on coupled processes, it
is convenient to use the Le Couteur-Newton relations \cite{LC} expressing
the $S$-matrix elements of all coupled processes in terms of the Jost matrix
determinant $d(k_1,k_2,k_3)$ that is the real analytic function with the only
square-root branch-points at $k_i=0$. Now we must find a proper uniformizing
variable for the 3-channel case. However, it is impossible to map the
8-sheeted Riemann surface onto a plane with the help of a simple function.
With the help of a simple mapping, a function, determined on the 8-sheeted
Riemann surface, can be uniformized only on torus. This is
unsatisfactory for our purpose. Therefore, we neglect the influence of the
$\pi\pi$-threshold branch point (however, unitarity on the $\pi\pi$-cut is
taken into account). An approximation like that means the consideration of
the nearest to the physical region semi-sheets of the Riemann surface. In
fact, we construct a 4-sheeted model of the initial Riemann surface
approximating it in accordance with our approach of a consistent account of
the nearest singularities on all the relevant sheets. The uniformizing
variable can be chosen as
\be \label{w}
w=\frac{k_2+k_3}{\sqrt{m_\eta^2-m_K^2}}.
\ee
It maps our model of the 8-sheeted Riemann surface onto the $w$-plane
divided into two parts by a unit circle centered at the origin.
\vspace{-1.mm}
\begin{figure}[ht]
%\vskip 3.cm
\centering
\epsfig{file=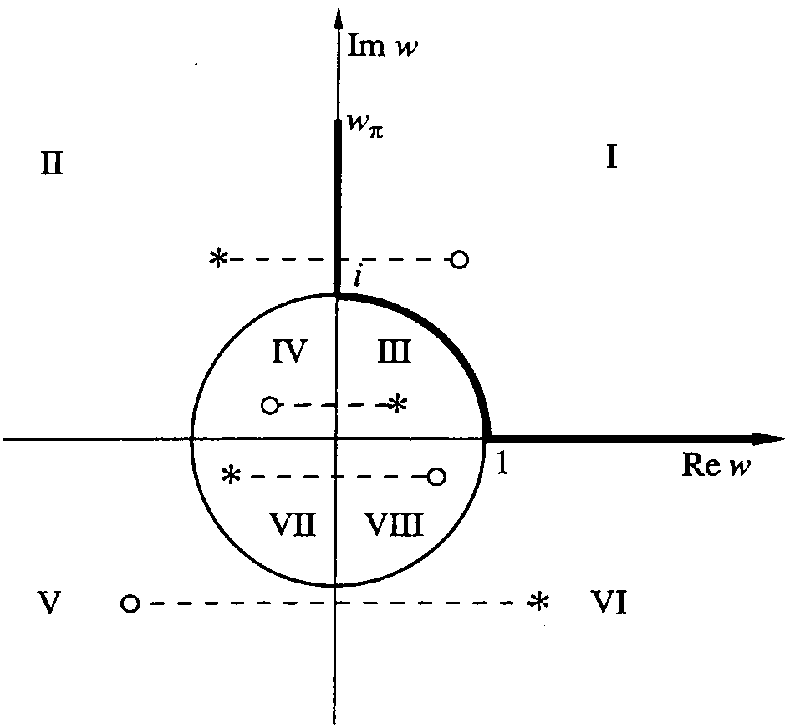, width=8cm}
\vskip -.05cm
\caption{
Uniformization plane for the \protect$\pi\pi$ scattering amplitude.
}
\label{fig:w.plane}
\end{figure}
On Fig.\ref{fig:w.plane}, the Roman numerals (I,II,\ldots,VIII) denote the
images of corresponding sheets of the Riemann surface; the thick line
represents the physical region (the points $w_\pi$, i and 1 are the
$\pi\pi$, $K\overline{K}$ and $\eta\eta$ thresholds, respectively). The
depicted positions of poles ($*$) and of zeros ($\circ$) give the
representation of the type ({\bf a}) resonance in $S_{11}$. The dashed lines
indicate a "pole-zero" symmetry required for elastic unitarity in the
($w_\pi$, i)--region.
 Here the left-hand cuts are neglected
in the Riemann-surface structure, and contributions on these cuts will be
taken into account in the background.

The Le Couteur-Newton relations are somewhat modified with taking
account of the used model of the Riemann surface (note that on the $w$-plane
the points $w_0$, $-w_0^{-1}$, $-w_0$, $w_0^{-1}$ correspond to the
$s$-variable point $s_0$ on sheets I, IV, V, VIII, respectively):
\bea \label{w:Le Couteur-Newton}
S_{11}=\frac{d^* (-w^*)}{d(w)},~~\qquad
S_{22}=\frac{d(-w^{-1})}{d(w)},~~\qquad S_{33}=\frac{d(w^{-1})}{d(w)},\\
D_{33}=\frac{d^*({w^*}^{-1})}{d(w)},\qquad D_{22}=\frac{d^*
(-{w^*}^{-1})}{d(w)},\qquad D_{11}=\frac{d(-w)}{d(w)}.\nn
\eea
Taking the $d$-function
as $d=d_B d_{res}$ where $d_B$, describing the background, is
$$d_B=\mbox{exp}[-i\sum_{n=1}^{3}k_n(\alpha_n+i\beta_n)],$$
moreover, the $\pi\pi$ background is taken to be elastic
up to the $K\overline{K}$ threshold.
The resonance part is
$$d_{res}(w)=w^{-\frac{M}{2}}\prod_{r=1}^{M}(w+w_{r}^*)$$
where $M$ is the number of resonance zeros.
%\newpage
\section{Combined analysis of experimental data}

We analyzed in a combined way the data on three processes
$\pi\pi\to\pi\pi,K\overline{K},\eta\eta$ in the channel with
$I^GJ^{PC}=0^+0^{++}$. For the $\pi\pi$-scattering, the data from the
threshold to 1.89 GeV are taken from work by B. Hyams et al.
\cite{Hyams}; below 1 GeV, from many works \cite{Hyams}. For $\pi\pi\to
K\overline{K}$, practically all the
accessible data are used \cite{Wetzel}. The $|S_{13}|^2$ data for
$\pi\pi\to\eta\eta$ from the threshold to 1.72 GeV are taken from ref.
\cite{Binon}.
As the data, we use
the results of phase analyses which are given for phase shifts of the
amplitudes and moduli of the $S$-matrix elements.

\begin{figure}[ht]
\centering
\begin{tabular}{cc}
\epsfxsize=8cm\epsffile{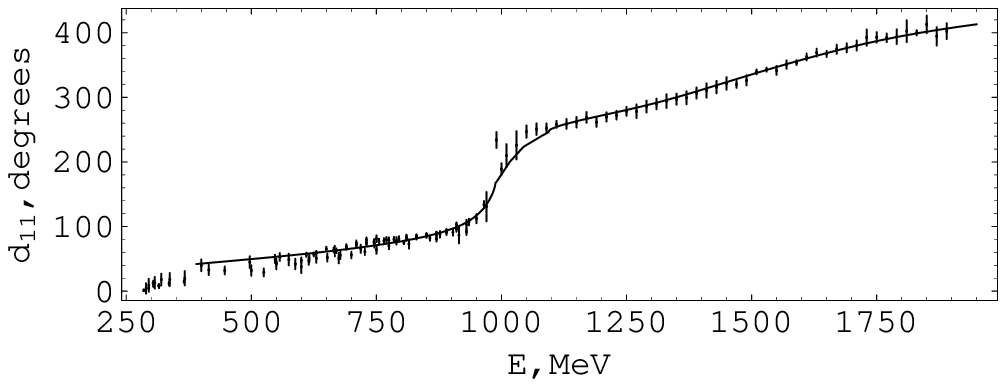}&
\epsfxsize=8cm\epsffile{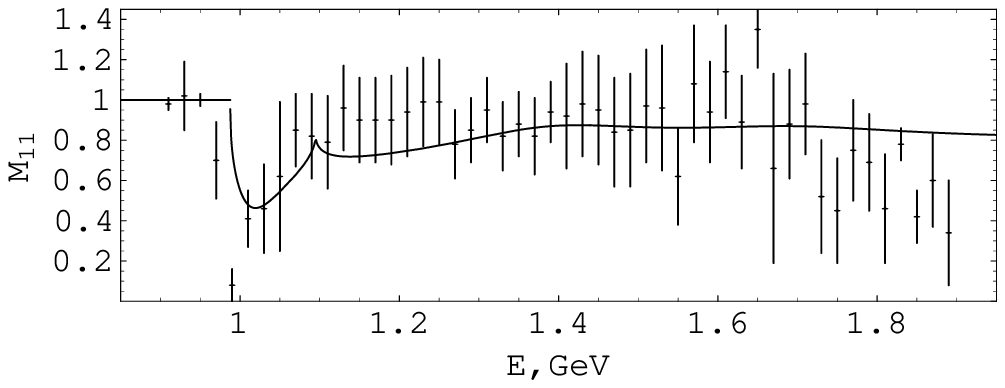}\\
\epsfxsize=8cm\epsffile{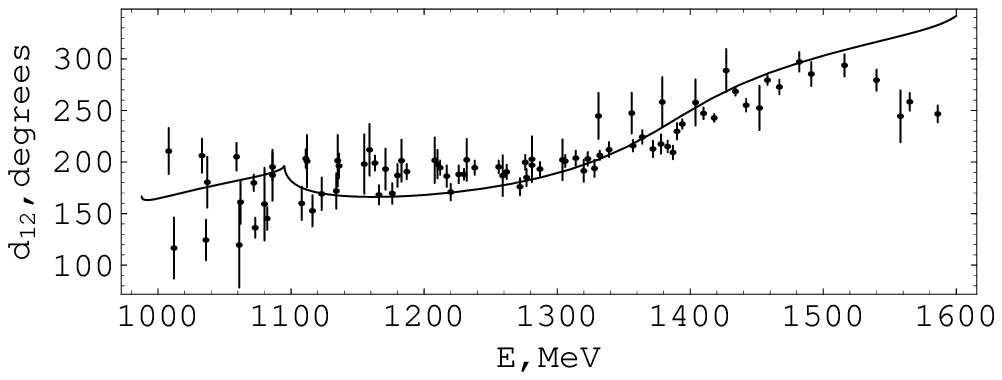}&
\epsfxsize=8cm\epsffile{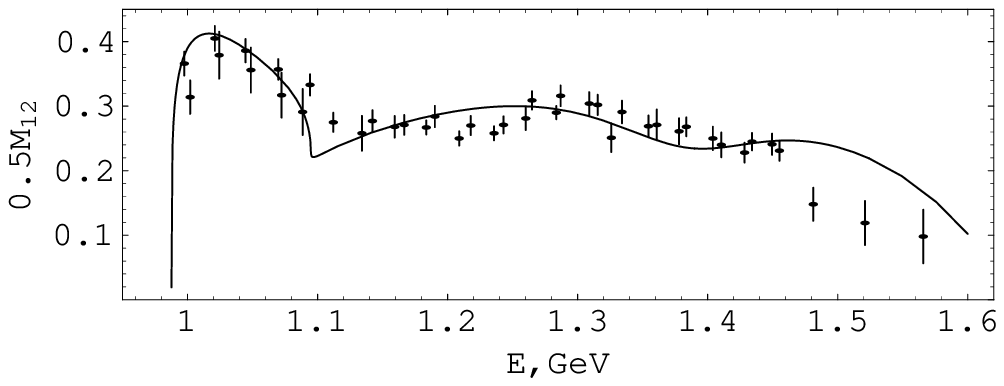}\\
\epsfxsize=8cm\epsffile{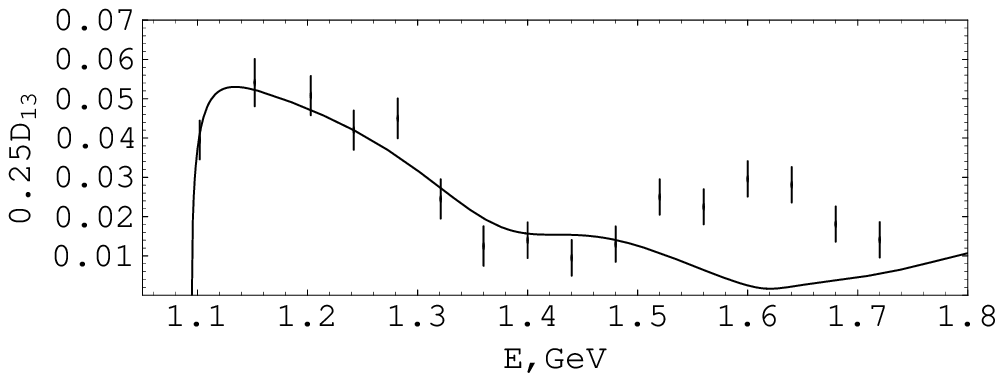}
\end{tabular}
\vskip -.1cm
\caption{Comparison of obtained energy dependences with experimental data.}
\label{fig:quantities}
\end{figure}

We obtain a satisfactory description: for the $\pi\pi$-scattering from
$\sim$ 0.4 GeV to 1.89 GeV ($\chi^2/\mbox{ndf}\approx 1.29$); for the
process $\pi\pi\to K\overline{K}$ from the threshold to $\sim$ 1.5 GeV
($\chi^2/\mbox{ndf}\approx 2.8$); for the $|S_{13}|^2$ data of the reaction
$\pi\pi\to\eta\eta$ from the threshold to 1.5 GeV ($\chi^2/\mbox{ndf}
\approx 0.95$). The total $\chi^2/\mbox{ndf}$ for all three processes
is 1.95; the number of adjusted parameters is 29. The background parameters
(in GeV$^{-1}$ units) are $\alpha_1=1.51, \beta_1=0.0482, \alpha_2=-0.93,
\beta_2=0.139,\beta_3=0.89$. On Figures \ref{fig:quantities}, we demonstrate
energy dependences of phase shifts and moduli of the matrix elements of
processes $\pi\pi\to \pi\pi,K\overline{K},\eta\eta$, compared with the
experimental data: $S_{ij}=M_{ij}e^{id_{ij}}$,
$D_{13}\equiv M_{13}^2$ ~($E=\sqrt{s}$).

\vspace{-0.1mm}

Let us indicate the obtained poles of clusters for resonances on the
complex energy plane $\sqrt{s}$, (in MeV units):~
for $f_0 (600)$, type ({\bf a}): $661\pm14-i(595\pm22)$ on sheet II,
$649\pm16-i(595\pm9)$ on sheet III, $611\pm17-i(595\pm28)$ on sheet VI,
and $623\pm15-i(595\pm26)$ on sheet VI;~
for $f_0 (980)$: $1006\pm5-i(34\pm8)$ on sheet II, and $962\pm18-
i(74\pm20)$ on sheet III;~
for $f_0 (1370)$, type ({\bf b}): $1384\pm21-i(50\pm29)$ on sheet III,
$1384\pm20-i(78\pm26)$ on sheet IV, $1384\pm20-i(182\pm23)$ on sheet V,
$1384\pm20-i(154\pm25)$ on sheet VI;
for $f_0 (1500)$, type ({\bf d}): $1505\pm22-i(320\pm25)$ on sheet II,
$1500\pm30-i(186\pm23)$ of the 2nd order on sheet III,
$1505\pm20-i(204\pm31)$ on sheet IV, $1505\pm21-i(320\pm30)$ on sheet V,
$1513\pm28-i(318\pm27)$ of the 2nd order on sheet VI, and
$1505\pm20-i(204\pm35)$ on sheet VII;~
for $f_0 (1710)$, type ({\bf c}): $1699\pm22-i(179\pm26)$ on sheet V,
$1699\pm22-i(168\pm23)$ on sheet VI, $1699\pm22-i(71\pm19)$ on sheet VII,
and $1699\pm26-i(82\pm18)$ on sheet VIII.
(To reduce the number of adjusted parameters, we supposed that simple poles
of the resonance clusters arise from the simplest 3-channel Breit-Wigner
form.)

Note a surprising result obtained for the $f_0(980)$ state.
It turns out that this state lies slightly above the $K\overline{K}$
threshold and is described by a pole on sheet II and by a shifted pole on
sheet III under the $\eta\eta$ threshold without an accompaniment of the
corresponding poles on sheets VI and VII, as it was expected for standard
clusters. This corresponds to the description of the $\eta\eta$ bound
state.

For now, we did not calculate coupling constants in the 3-channel
approach, because here rather much variants of combinations of the resonance
cluster types are possible and not all the ones are considered to choose the
better variant, though already a satisfactory description is obtained.
Therefore, for subsequent conclusions, let us mention the results for
coupling constants from our previous 2-channel analysis, which have been
calculated through the residues of the amplitudes at the poles on the
relevant sheets.
\begin{center} Table 3: Coupling constants of the $f_0$ states with $\pi\pi$
($g_1$) and $K\overline{K}$ ($g_2$) systems.
%\end{center}
%\vskip0.3truecm
\begin{tabular}{|c|c|c|c|c|} \hline {}  &
$f_0(665)$ & $f_0(980)$ &
$f_0(1370)$ & $f_0(1500)$\\ \hline $g_{1}$, GeV
& ~$0.652\pm 0.065$~ &
~$0.167 \pm 0.05$~~ & ~$0.116 \pm 0.03$~ &
~$0.657 \pm 0.113$~\\ \hline
$g_2$, GeV & ~$0.724\pm0.1$~~~~ &
~$0.445\pm0.031$~ & ~~$0.99\pm0.05$~ &
~$0.666\pm0.15$~~\\ \hline \end{tabular}\end{center}
\bigskip

On the basis of the types of pole clusters of the considered resonances and
taking into account that (as it is seen from Table 3) the $f_0(980)$ and
especially the ${f_0}(1370)$ are coupled essentially more strongly to the
$K\overline{K}$ system than to the $\pi\pi$ one, we have concluded that these
states have a dominant $s{\bar s}$ component. The $f_0(1500)$ has the
approximately equal coupling constants with the $\pi\pi$ and $K\overline{K}$
systems, which apparently could point up to its dominant glueball component.
In the 2-considerations, $f_0 (1710)$ is represented by the pole cluster
corresponding to a state with the dominant $s{\bar s}$ component.

Our 3-channel conclusions on the basis of resonance cluster types
generally confirm the ones drawn in the 2-channel analysis,
besides the above surprising conclusion about the $f_0(980)$ nature.

Masses and widths of these states should be calculated from the pole
positions. If we take the resonance part of amplitude as
$$T^{res}=\frac{\sqrt{s}\Gamma_{el}}{m_{res}^2-s-i\sqrt{s}\Gamma_{tot}},$$
we obtain for masses and total widths the following values (in MeV units):
for $f_0 (665)$, 889 and 1190;
for $f_0 (980)$, 1006 and 64;
for $f_0 (1370)$, 1386 and 156;
for $f_0 (1500)$, 1539 and 640;
for $f_0 (1710)$, 1701 and 164.

%\newpage
\section{Lower scalar $0^{++}$ nonets}

Although at present many states have been discovered in the scalar mesonic
sector \cite{PDG-02}, however, their assignment to quark-model configurations is
problematic -- one can compare various variants of that assignment,
for example, \cite{Lanik}-\cite{Volkov-Yudich}.

On the basis of obtained results, we can propose a following assignment of
scalar mesons below 1.9 GeV to lower nonets. First of all, we exclude from
this consideration the $f_0(980)$ as the $\eta\eta$ bound state. Then
we propose to include to the lowest nonet the isovector $a_0(980)$, the
isodoublet $K_0^*(900)$ (or $\kappa(800)$), and two isoscalars $f_0(600)$
and $f_0(1370)$ as mixtures of the eighth component of octet and the SU(3)
singlet.  Note also that we consider the $K_0^*(900)$ (or $\kappa$),
observed at analysing the $K-\pi$ scattering \cite{Ishida} and at studying
the decay $D^+\to K^-\pi^+\pi^+$ (Fermilab experiment E791) \cite{Gobel}.
Then the Gell-Mann--Okubo formula
\be \label{GM-O}
3m_{f_8}^2=4m_{K^*}^2-m_{a_0}^2
\ee
gives $m_{f_8}=0.87$ GeV. Our result for the $\sigma$-meson mass is
$m_\sigma\approx0.889\pm0.02$ GeV (if $m_\kappa=0.8$, $m_{f_8}\approx0.73$).

The second relation for masses of nonet, which is obtained only on basis of
the quark contents of the nonet members and somehow restricts mass of the
SU(3) singlet, is
\be \label{8-0-1/2}
m_\sigma+m_{f_0(1370)}=2m_\kappa.
\ee
The left-hand side of this relation is $\sim$ 25 \% bigger than the
right-hand one if to take our mass values.

The next nonet could be formed of the isovector $a_0(1450)$, the isodoublet
$K_0^*(1430-1450)$, and of the $f_0(1500)$ and $f_0(1710)$ as mixtures of
the eighth component of octet (mixed with a glueball) and the SU(3) singlet.
From the Gell-Mann--Okubo formula we obtain $m_{f_8}\approx1.45$ GeV.
In second formula
\be \label{r:8-0-1/2}
m_{f_0(1500)}+m_{f_0(1710)}=2m_{K^*(1450)},
\ee
the left-hand side is $\sim$ 12 \% bigger than the right-hand one.

Though the Gell-Mann--Okubo formula is fulfilled for both nonets rather
satisfactorily, however, the breaking of 2nd relation (especially
for the lowest nonet) tells us that the $\sigma-f_0(1370)$ and
$f_0(1500)-f_0(1710)$ systems get additional contributions absent in
the $K_0^*(900)$ and $K_0^*(1450)$, respectively.

%\newpage
\section{Conclusions}
\begin{itemize}
\item
In a combined model-independent analysis of data on the $\pi\pi$
scattering from 0.4 to 1.89 GeV and processes $\pi\pi\to K\overline{K},
\eta\eta$ from the thresholds to 1.5 GeV, a confirmation of the
$\sigma$-meson with mass 0.889 GeV is obtained once more. This mass value
rather accords with prediction ($m_\sigma=m_\rho$) on the basis of mended
symmetry by Weinberg \cite{Weinberg}.
\item
Consideration of the $\eta\eta$ channel is necessary for a consistent
and reasonable representation of the obtained resonances. For satisfactory
description of the processes $\pi\pi\to K\overline{K},\eta\eta$ above
$\sim$ 1.5 GeV, an allowance for channels (first of all, the $\eta\eta'$
one) opening in this region is necessary.
\item
The $f_0(980)$, ${f_0}(1370)$ and $f_0 (1710)$ have the
dominant $s{\bar s}$ component. Moreover, we obtain an additional indication
for $f_0(980)$ to be the $\eta\eta$ bound state. Remembering a dispute
\cite{Zou} whether the $f_0(980)$ is narrow or not, we agree rather with the
former. Of course, it is necessary to make analysis of other relevant
processes, first of all, $J/\psi$ and $\phi$ decays. Note also that our
conclusion about the large $s{\bar s}$ component
for the $f_0 (1710)$ quite is consistent with the experimental facts that
this state is observed in $\gamma\gamma\to K_SK_S$ \cite{Braccini} and not
observed in $\gamma\gamma\to\pi^+\pi^-$ \cite{Barate}.
\item
As to the $f_0(1500)$, we suppose that it is practically the eighth
component of octet mixed with a glueball being dominant in this state.
\item
An assignment of the scalar mesons below 1.9 GeV to lower nonets is
proposed. Note that this assignment moves a number of questions and does
not put the new ones. It is clear that now an adeguate mixing scheme should
be found.
\end{itemize}

\noindent  ACKNOWLEDGEMENTS -
This work has been supported by the Grant Program of Plenipotentiary of
Slovak Republic at JINR. M.N. were supported in part by the Slovak
Scientific Grant Agency, Grant VEGA No. 2/3105/23; and D.K., by Grant
VEGA No. 2/5085/99.

\end{document}